\documentclass[aps,prl,twocolumn,superscriptaddress,showpacs]{revtex4-1}
\usepackage{amsmath}
\usepackage{amssymb}
\usepackage{graphicx}
\usepackage{dcolumn}
\usepackage{bm}

\begin{document}

\title{Spectral properties of Dirac billiards at the van Hove singularities}

\author{B.~Dietz}
\email{dietz@ikp.tu-darmstadt.de}
\affiliation{Institut f{\"u}r Kernphysik, Technische Universit{\"a}t Darmstadt, 
Schlossgartenstr. 9, D-64289 Darmstadt, Germany}

\author{T.~Klaus}
\affiliation{Institut f{\"u}r Kernphysik, Technische Universit{\"a}t Darmstadt, 
Schlossgartenstr. 9, D-64289 Darmstadt, Germany}

\author{M.~Miski-Oglu}
\altaffiliation{Present address: GSI Helmholtzzentrum f\"ur Schwerionenforschung GmbH,
Planckstr. 1, D-64291 Darmstadt, Germany}
\affiliation{Institut f{\"u}r Kernphysik, Technische Universit{\"a}t Darmstadt, 
Schlossgartenstr. 9, D-64289 Darmstadt, Germany}

\author{A.~Richter}
\email{richter@ikp.tu-darmstadt.de}
\affiliation{Institut f{\"u}r Kernphysik, Technische Universit{\"a}t Darmstadt, 
Schlossgartenstr. 9, D-64289 Darmstadt, Germany}

\author{M.~Wunderle}
\affiliation{Institut f{\"u}r Kernphysik, Technische Universit{\"a}t Darmstadt, 
Schlossgartenstr. 9, D-64289 Darmstadt, Germany}

\author{C. Bouazza}
\altaffiliation{Present address: Coll\'ege de France, 11 Place Marcelin-Berthelot, 75231 Paris Cedex 05, France}
\affiliation{Institut f{\"u}r Kernphysik, Technische Universit{\"a}t Darmstadt,
Schlossgartenstr. 9, D-64289 Darmstadt, Germany}

\date{\today}

\begin{abstract}
We study distributions of the ratios of level spacings of a rectangular and an Africa-shaped superconducting microwave resonator containing circular scatterers on a triangular grid, so-called Dirac billiards (DBs). The high-precision measurements allowed the determination of \emph{all} 1651 and 1823 eigenfrequencies in the first two bands, respectively. The resonance densities are similar to that of graphene. They exhibit two sharp peaks at the van Hove singularities, that separate the band structure into regions with a linear and a quadratic dispersion relation, respectively. In their vicinity we observe rapid changes, e.g., in the wavefunction structure. Accordingly, the question arose, whether there the spectral properties are still determined by the shapes of the DBs. The commonly used statistical measures, however, are no longer applicable whereas, as demonstrated in this Letter, the ratio distributions provide most suitable ones.

\end{abstract}


\maketitle

{\it Introduction.}--- The focus of the experiments with microwave photonic crystals~\cite{Yablonovitch1989} reported in this Letter were the spectral properties of finite-size graphene sheets~\cite{Beenakker2008,Castro2009}, particularly in the vicinity of the van Hove singularities~\cite{VanHove1953} (vHSs) exhibited by the density of states (DOS)~\cite{Wallace1947,Nierenberg1951,Castro2009}. Graphene, a monoatomic layer of carbon atoms arranged on a honeycomb lattice, has exceptional electronic properties that stem from the shapes of its conduction and its valence band. They touch each other conically at the six corners of the hexagonal Brillouin zone constituted by two independent Dirac points (DPs) denoted by {\bf K}$_\pm$. Previous studies focused on the energy region around the DPs, where the dispersion relation is linear and thus, graphene exhibits \emph{relativistic} phenomena~\cite{Geim2007,Avouris2007,Beenakker2008,Castro2009,Abergel2010}. The band structure, however, becomes more complex with increasing energy and eventually looses its linearity and becomes quadratic~\cite{Dietz2015}. The transition takes place at its saddle points, the {\bf M} points, corresponding to the vHSs in the DOS. They generally occur in two-dimensional crystals with a periodic structure~\cite{VanHove1953,Cappelluti1996,Gonzalez2000,Vozmediano2002,Wilder1998,Kuerti2002,Yan2014,Havener2014,McChesney2010,Mak2011} and give rise to logarithmic divergences in the DOS. As a consequence, arbitrarily weak interactions can produce large effects in the electronic behavior of graphene.  Once the Fermi energy approaches a vHS, as, e.g., in hole-doped cuprates, the presence of the singularity may lead to an enhancement of ferromagnetism, antiferromagnetism or superconductivity~\cite{Ziitko2009,Hirsch1986,Pattnaik1992,Markiewicz1994,Hlubina1997,Ziletti2015}. The vHS, actually, is a topoligical, critical point where a quantum Lifshitz phase transition takes place~\cite{Vidhyadhiraja2009,Son2011,Chen2011a,Gradinar2012,Bellec2013,Dietz2013,Xu2015}. 

Tight-binding model (TBM) calculations revealed, that similarly the wavefunction structure varies rapidly in the vicinity of the vHSs. Consequently, it is not obvious that there the spectral properties only depend on the shape of the DB, like they do close to the DPs and the band edges~\cite{Dietz2015}. A prereqisite for the applicability of commonly used statistical measures for their investigation is the unfolding, i.e., the rescaling of the levels to mean spacing one. This is not possible close to the VHSs. The main objective of the present Letter is thus to experimentally study there the spectral properties of graphene flakes (dots) in terms of the dimensionless ratios of level spacings~\cite{Oganesyan2007,Atas2013}. This, hitherto, was not possible, because of the, due to the high level density, required precision of the measurements. Here, we exploited the fact, that the peculiar shape of the first two bands of graphene arises due to the symmetry properties of its honeycomb structure which is formed by two interpenetrating triangular lattices with threefold symmetry. It, actually, has been reproduced using two-dimensional electron gases, molecular assemblies, ultracold atoms~\cite{Singha2011,Nadvornik2012,Gomes2012,Tarruell2012,Uehlinger2013} photonic crystals~\cite{Bittner2010,Kuhl2010,Sadurni2010,Bittner2012,Bellec2013,Rechtsman2013,Rechtsman2013a,Khanikaev2013} or, generally, systems referred to as artificial graphene~\cite{Polini2013a}. Furthermore, experimental studies have been performed with graphene quantum dots, so-called graphene billiards~\cite{Miao2007,Ponomarenko2008,Westervelt2008,Guettinger2010,Guettinger2012}. We performed high-resolution measurements with superconducting, macroscopic-size microwave Dirac billiards (DBs)~\cite{Bittner2012,Dietz2015} with the shapes of a rectangle and the African continent~\cite{Berry1987,Huang2010,Huang2011}. In~\cite{Dietz2015} we analyzed the spectral properties of graphene near the {\bf K}$_\pm$ points and around the center of the first Brillouin zone, the $\boldsymbol\Gamma$ points at the band edges. The experiments were motivated furthermore by the inconsistent experimental~\cite{Ponomarenko2008} and numerical~\cite{Libisch2009,Huang2010,Huang2011,Wurm2009,Wurm2011,Ni2012} results for the spectral properties of Africa-shaped graphene billiards. 

\begin{figure}[ht!]
{\includegraphics[width=\linewidth]{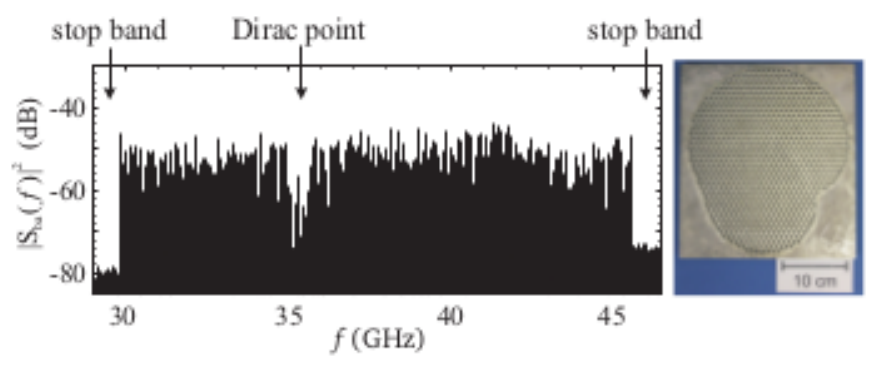}}
\caption{Transmission spectrum (l.h.s.) in the frequency range of the first and the second band of the Dirac billiard shown in the photograph with the lid removed (r.h.s.). It has the shape of Africa and contains $\approx 900$ metal cylinders arranged on a triangular lattice. A region of low resonance density around the DP is clearly visible.} 
\label{fig1}
\end{figure}

\begin{figure}[t!]
{\includegraphics[width=\linewidth]{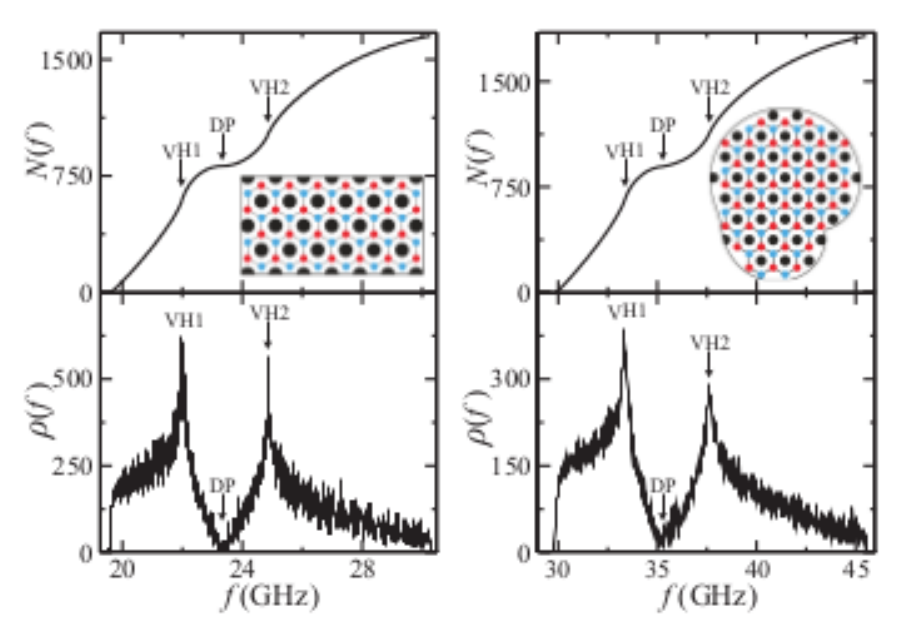}}
\caption{Integrated resonance density (upper panels) and resonance density (lower panels) of the rectangular (left part) and the Africa-shaped (right part) DBs. The insets show a schematic view of the corresponding DB (black). The red (gray) and blue (dark gray) dots mark the voids between the metal cylinders of the photonic crystal situated at the sites of the two interpenetrating triangular lattices.} 
\label{fig2}
\end{figure}

{\it Experimental setup and spectral properties.}--- A photograph of the Africa-shaped microwave DB is shown on the r.h.s.~of Fig.~\ref{fig1}. 
The DBs consisted of a basin and a lid made from brass plates. The basin contained a photonic crystal which was constructed by milling $\approx 900$ metal cylinders arranged on a triangular grid out of a brass plate. The lattice constant, i.e., the distance between neighboring cylinders, was $a_L=12$~mm and $a_L=8$~mm, respectively, and the radius of the cylinders was $R=a_L/4$. To attain superconductivity at liquid helium temperature, i.e., at $T_{\rm LHe}=4.2$~K, the lid and the basin were coated with lead, which has a critical temperature $T_c=7.2$~K. The height of the resonators was $d=3$~mm and the range of excitation frequencies $f$ of the microwaves that were coupled into the resonator was chosen as $0\leq f\lesssim 50$~GHz. Up to the maximal frequency the electric field modes are described by the scalar Helmholtz equation with Dirichlet boundary conditions at the walls of the basin and the cylinders. This equation is mathematically equivalent to the Schr\"odinger equation of a quantum billiard of the same shape containing circular scatterers at the positions of the latter~\cite{Stoeckmann1990,Richter1999}. The honeycomb structure of the DBs is generated by the voids at the centers of the triangles formed by, respectively, three of the cylinders; see insets of Fig.~\ref{fig2}. The corresponding \emph{empty} quantum billiard (QB) is obtained by removing the cylinders~\cite{Stoeckmann1990,Richter1999}. We chose DBs with the shapes of a rectangle and of Africa because their classical dynamics is fully integrable and fully chaotic with no nongeneric contributions from bouncing-ball orbits~\cite{Sieber1993}, respectively. The choice of the latter shape actually was motivated by the seminal work on neutrino billiards~\cite{Berry1987}. 

Figure~\ref{fig1} shows a transmission spectrum of the Africa-shaped DB. Below the frequency of the lower band edge at $f_{\rm lBE}=29.84$~GHz no resonances were detected, and above the upper one at $f_{\rm uBE}=45.52$~GHz a second broad band gap is observed. In between a narrow gap of low resonance density is clearly visible, which separates the first and the second band. It is situated around the frequency $f_{\rm D}=35.32$~GHz of the DP. The corresponding frequency values for the rectangular DB are provided in Ref.~\cite{Dietz2015}. The positions of the resonances yield the eigenfrequencies $f_i$ of the microwave DBs. Because of the high quality factor $Q\gtrsim 5\times 10^5$ of the superconducting microwave resonators, \emph{complete} sequences of altogether $1651$ and $1823$ eigenfrequencies could be identified in the first two bands of the rectangular and the Africa DB, respectively. Figure~\ref{fig2} shows their integrated density (upper panels) and their density of the eigenfrequencies (lower panels). The latter corresponds to the DOS per unit cell~\cite{Wallace1947,Nierenberg1951,Castro2009}. 
It ressembles that of graphene~\cite{Wallace1947,Hobson1953,Castro2009} with a minimum at the Dirac frequency. The two sharp peaks at the frequencies of the {\bf M} points evolve into logarithmic vHSs with increasing size of the sheet. The DOS of the Africa DB, in addition, exhibits a slight bump above the Dirac frequency, that is, an accumulation of eigenfrequencies. They correspond to edge states that are localized at the zigzag edges formed by the void structure~\cite{Wurm2011,Kuhl2010}. 

\begin{figure}[ht!]
{\includegraphics[width=\linewidth]{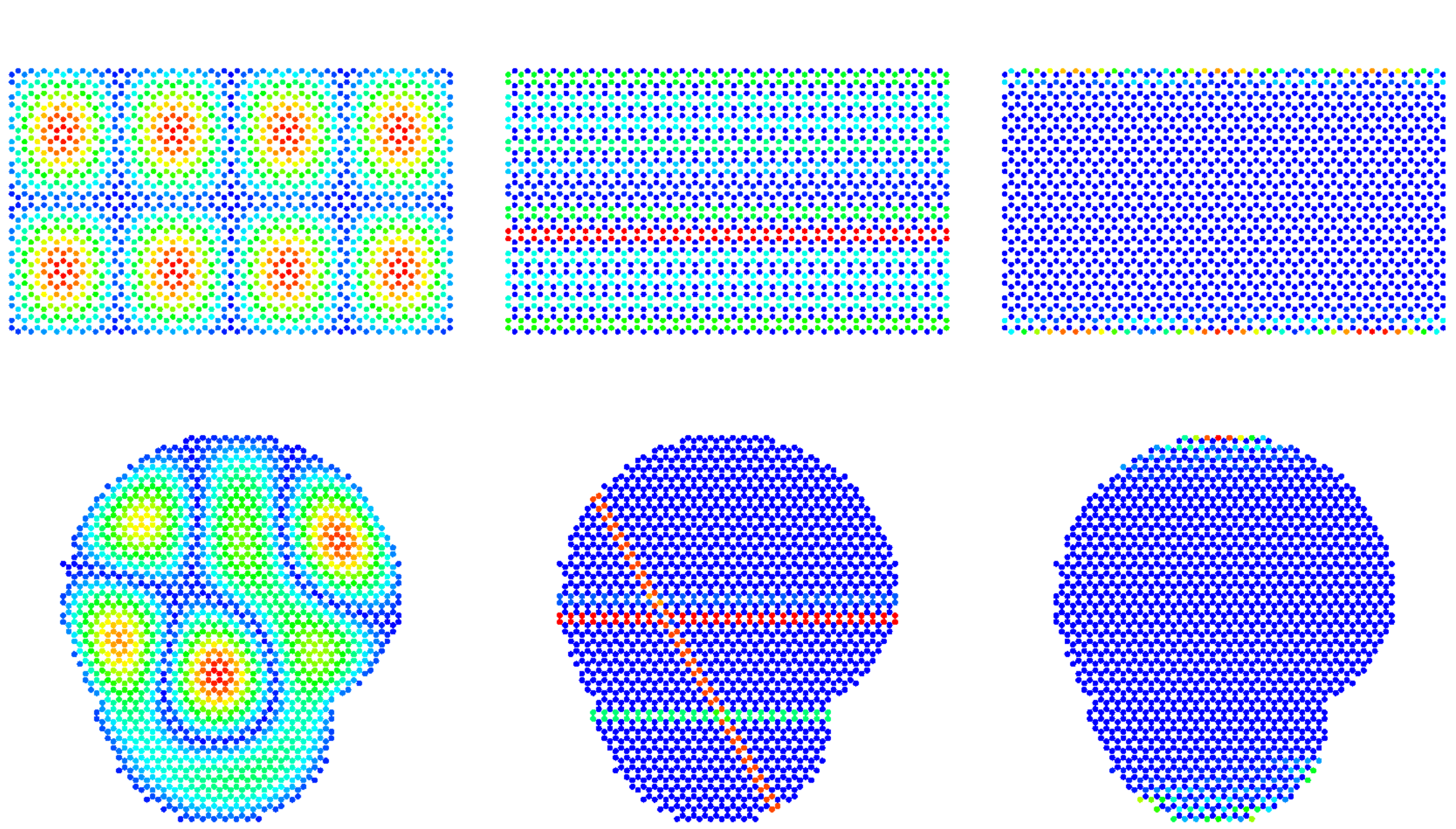}}
\caption{Computed intensity distributions of the wavefunctions of graphene sheets with the same honeycomb structure as the one formed by the voids (see insets of Fig.~\ref{fig2}) inside the rectangular (upper row) and Africa-shaped DB (lower row), close to a band edge (left panels), a vHS (middle panels) and the DP (right panels). Here, red (gray) and blue (dark gray) correspond to a maximal and a vanishing intensity, respectively.}
\label{fig2b}
\end{figure}

In the regions close to the band edges, the eigenfrequencies of the DBs are directly related to the eigenvalues of the associated QB~\cite{Dietz2015}. These observations were corroborated by TBM~\cite{Wallace1947,Reich2002} calculations for the honeycomb lattice formed by the voids inside the DBs. In Fig.~\ref{fig2b} we show for each DB a computed wavefunction intensity distribution in the vicinity of a band edge (left panels). They are identical with those of the QB. 
This similarity holds for the first $\approx 250$ and $\approx 90$ eigenstates of the rectangular and the Africa-shaped DB counted from the band edges, respectively. For the latter the number is smaller due to the deviation of its curved shape from that of the honeycomb lattice fitted into it. In the region around the DP, we found a linear interdependence between the eigenfrequencies of the DBs and the eigenvalues of the graphene billiard~\cite{Wurm2011} of corresponding shape. The associated wavefunctions shown in the right panels of Fig.~\ref{fig2b} exhibit no structure. This is in accordance with the low-energy approximation for graphene which implies a diverging effective wavelength. The intensity, actually, is nonvanishing only at the zigzag edges~\cite{Wurm2011}. 

We also investigated the spectral fluctuation properties of the frequencies $e_i=\vert f_i-f_0\vert$ with respect to the band edges, $f_0=f_{\rm lBE},\, f_{\rm uBE}$ and to the DP, $f_0=f_{\rm D}$~\cite{Mehta1990,Dietz2015}. For this, we unfolded the $e_i$ by replacing them by the smooth part of the integrated DOS, $\tilde e_i=N_{\rm smooth}(e_i)$. This procedure requires an analytical expression for $N_{\rm smooth}(e_i)$. For QBs it is given by Weyl's law~\cite{Weyl1912}. It, however, is not available for the DBs (see the upper panels of Fig.~\ref{fig2}). Therefore, we studied the spectral fluctuation properties for short sequences of 100 eigenfrequencies and obtained $N_{\rm smooth}(e_i)$ in terms of the polynomial best fitting the experimental $N(e_i)$. We came to the conclusion that both, in the regions around the $\boldsymbol\Gamma$ points and the DPs, the spectral properties coincide with those of the corresponding QB. The latter are only determined by the shape of the billiard, and in accordance with the Bohigas-Giannoni-Schmit conjecture~\cite{McDonald1979,Casati1980,Berry1981,Bohigas1984} for integrable and for time-reversal invariant chaotic systems. For the Africa-shaped DB we had to exclude the eigenfrequencies of the edge states; see Fig.~\ref{fig1}. A drawback of these statistical measures is that they require an unfolding of the eigenvalues. Consequently, they cannot be used to analyze the spectral properties in the vicinity of the vHSs. 

{\it Ratio distributions.}--- About ten years ago, a new statistical measure was proposed by Oganesyan and Huse~\cite{Oganesyan2007} which characterizes the correlations between consecutive spacings of adjacent eigenvalues or in our case, eigenfrequencies of the DB. The authors considered the distribution $P(\tilde r)$ of $\tilde r_i=\min\left(r_i,1/r_i\right)$ with $r_i=(e_{i+1}-e_i)/(e_i-e_{i-1})$. Recently, the distribution of the ratios $r_i$ and the $k$th overlapping ratio distribution $P(r^k)$ of spacings between the $k$th nearest neighbors $r_i^k=(e_{i+k+1}-e_i)/(e_{i+k}-e_{i-1})$ were introduced~\cite{Atas2013,Atas2013a}. In Ref.~\cite{Atas2013} a Wigner-like~\cite{Mehta1990} approximation was derived for the ratio distribution of the Gaussian orthogonal ensemble (GOE))~\cite{Mehta1990,Weidenmueller2009,Mitchell2010}, $P^{\rm GOE}(r)\simeq 27/8\cdot (r+r^2)/(1+r+r^2)^{5/2}$, that for Poissonian random numbers~\cite{Oganesyan2007} reads $P^{\rm Poisson}(r)=1/(1+r)^2$. Here, the GOE and Poisson statistics describe the spectral properties of generic chaotic and integrable systems~\cite{McDonald1979,Casati1980,Berry1981,Bohigas1984}, respectively. The accuracy was further improved and analytical expressions were given for the $k$th overlapping ratio distribution in Ref.~\cite{Atas2013a}. In Ref.~\cite{Chavda2014} the transition from Poissonian to GOE statistics was investigated thoroughly for the ratio distributions $P(r)$ and $P(\tilde r)$ in terms of the averages $\langle r\rangle$ and $\langle\tilde r\rangle$. The values for the limiting cases are $\langle r\rangle_{\rm Poi}=\infty ,\, \langle\tilde r\rangle_{\rm Poi}=0.39$ and $\langle r\rangle_{\rm GOE}=1.75 ,\, \langle\tilde r\rangle_{\rm GOE}=0.54$. The authors came to the conclusion that the critical values for the transition from Poisson to GOE are $\langle r\rangle_{\rm crit}=2.0 ,\, \langle\tilde r\rangle_{\rm crit}=0.5$. 

The quantities $\tilde r_i$, $r_i$ and $r_i^k$ are dimensionless. Therefore, as long as the DOS does not vary on the scale of the average spacing, no unfolding is needed~\cite{Oganesyan2007,Kollath2010,Collura2012,Chavda2013,Chavda2014}. An objective of the present Letter was to test whether their distributions provide a statistical measure which is sensitive to the behavior of the classical dynamics. Here, the vicinity of the sharp peaks in the DOS was of particular interest, since there the DOS, and thus the average spacing, varies rapidly. Actually, the latter becomes extremely small, so that the applicability of the ratio distributions becomes questionable. We demonstrate that they are employable in our finite-size systems. In addition, our aim was to give an answer to the question, whether there the spectral properties are also determined by the shape of the DB, using our highly precise experimental data. 

\begin{figure}[ht!]
{\includegraphics[width=\linewidth]{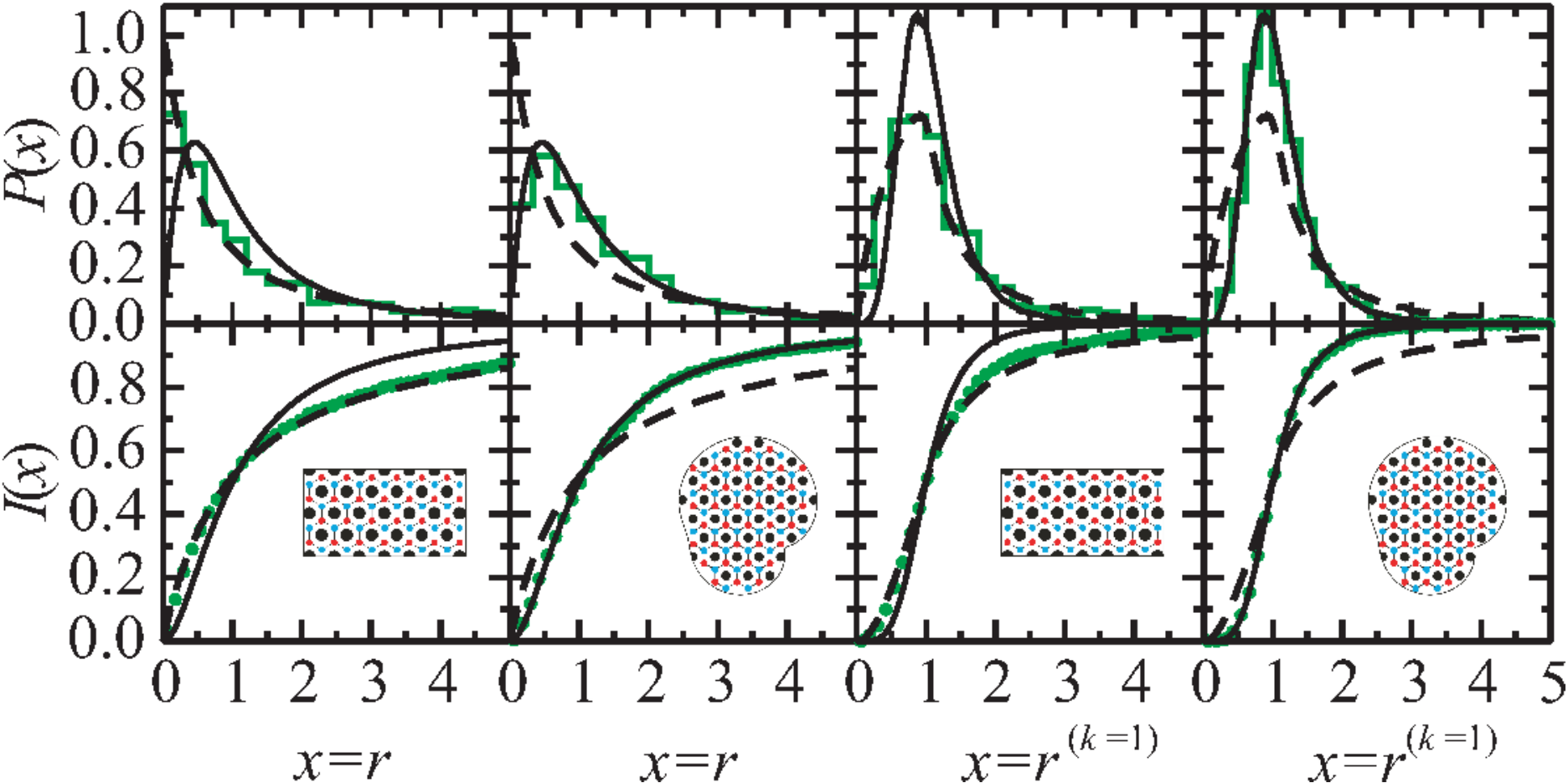}}
\caption{Left two columns: Ratio distributions (upper panels, green histograms) and their cumulative distributions (lower panels, green dots) for all eigenfrequencies in the first two bands of the rectangular and the Africa-shaped DB (see insets), in comparison to those for a Poissonian statistics (dashed lines) and the GOE result (full line). Right two columns: Same as the left ones for the ($k$=1)-overlapping ratio distributions.}
\label{fig3}
\end{figure}

In Fig.~\ref{fig3} all eigenfrequencies of the DBs shown in the insets of the lower panels were taken into account. In the upper panels the experimental (green histograms) ratio distributions (left two panels) and the ($k$=1)-overlapping ratio distributions (right two panels) are compared with those for the GOE (full lines) and for a Poisson process (dashed lines). In order to avoid the dependence on the bin sizes used for the histograms, we also evaluated the corresponding cumulative distributions $I(r)=\int_0^r{\rm d}r^\prime P(r^\prime)$, shown in the lower panels (green dots). The distributions coincide with those of a Poisson process for the rectangular DB, and with the GOE statistics for the Africa-shaped one, in accordance with the findings for the fluctuation properties of the unfolded eigenfrequencies. The agreement, actually, is so good, that the curves lie on top of each other.
The average values, $\langle r\rangle_{\rm rect}=2.70 ,\, \langle\tilde r\rangle_{\rm rect}=0.40$ and $\langle r\rangle_{\rm africa}=1.75 ,\, \langle\tilde r\rangle_{\rm africa}=0.53$ are close to those of the corresponding theoretical distribution. 

\begin{figure}[ht!]
{\includegraphics[width=\linewidth]{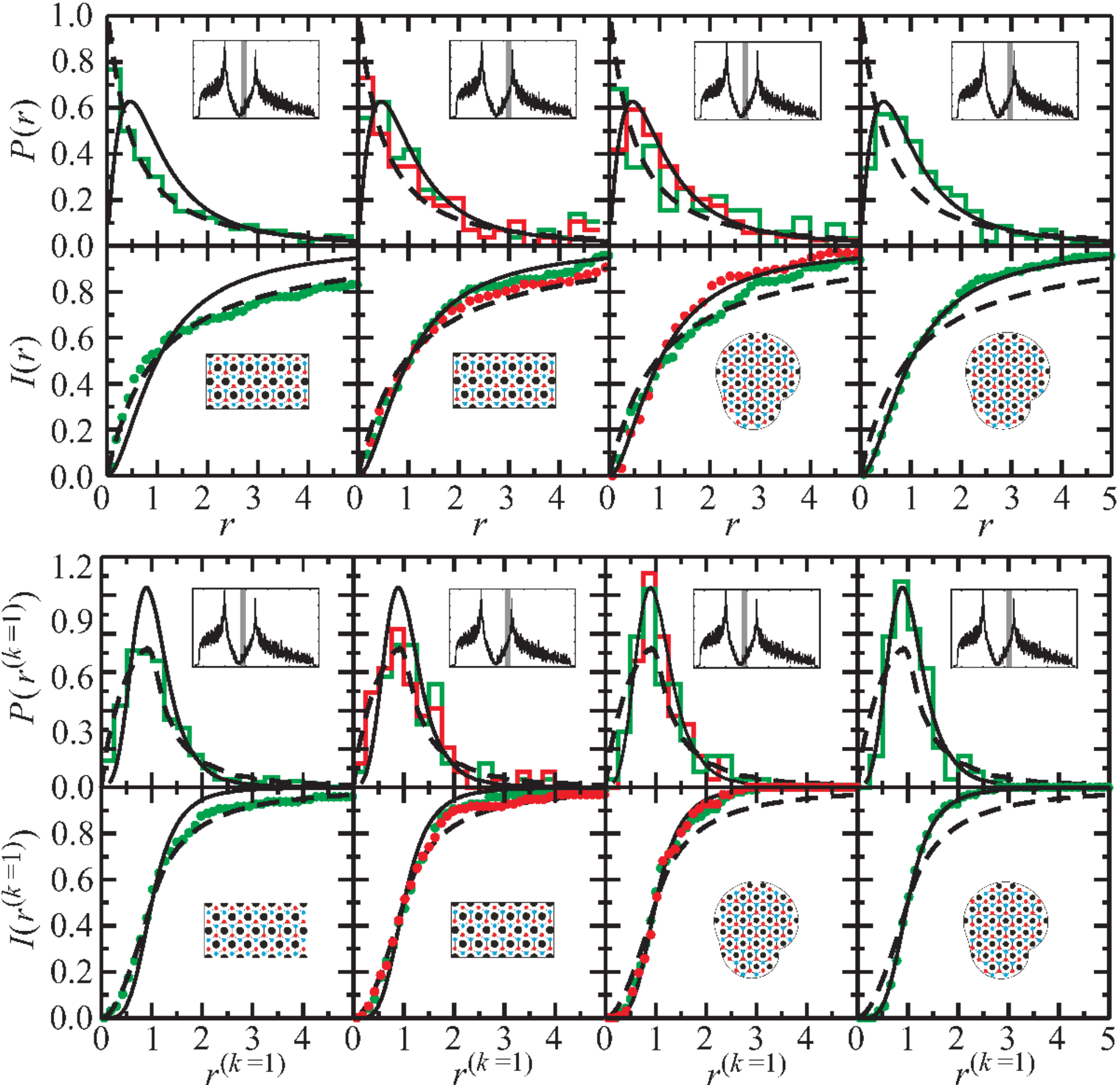}}
\caption{Upper part: Experimental ratio distributions (upper panels, green histograms) and their cumulative distributions (lower panels, green dots) above the Dirac frequency and below the upper vHS for the rectangular DB and the Africa-shaped one (see insets) in comparison to those for a Poissonian statistics (dashed lines) and the GOE result (full line). The red histograms show distributions after extracting nongeneric contributions. Lower part: Same as upper part for the ($k$=1)-overlapping ratio distributions.}
\label{fig4}
\end{figure}

We also investigated the statistical properties of the ratios separately for sequences of 100 eigenfrequencies close to the band edges, around the Dirac frequency and near the vHSs. At the band edges, the experimental ratio distributions and the ($k$=1)-overlapping ratio distributions of the rectangular and the Africa-shaped DB agree well with the Poisson and the GOE curves, respectively. The same holds for the averages, $\langle r\rangle_{\rm rect}=2.63 ,\, \langle\tilde r\rangle_{\rm rect}=0.41$ and $\langle r\rangle_{\rm africa}=1.77 ,\, \langle\tilde r\rangle_{\rm africa}=0.52$. Figure~\ref{fig4} shows the experimental (green histograms and dots) ratio distributions (upper part) and the ($k$=1)-overlapping ratio distributions (lower part) in regions above the DP and below the upper vHS; see insets. In the Dirac region we omitted the first 10 eigenfrequencies below and above the DP, since they yield nongeneric contributions as can be deduced from the features of the intensity distributions; see right panels of Fig.~\ref{fig2b}. We obtained for the rectangular DB and, interestingly, also for the Africa-shaped one a good agreement with the Poissonian statistcs. In accordance with these observations,  $\langle r\rangle_{\rm rect}=2.41 ,\, \langle\tilde r\rangle_{\rm rect}=0.41$ and $\langle r\rangle_{\rm africa}=2.08 ,\, \langle\tilde r\rangle_{\rm africa}=0.46$. For the latter, the deviation from the expected GOE behavior is attributed to the edge states present above the DP, see Fig.~\ref{fig2}. Their intensity distributions are localized at the zigzag edges and thus exhibit a nongeneric behavior. Only after their omission, a very good agreement with the GOE is obtained (red histograms and dots). Then, the average ratios equal $\langle r\rangle_{\rm africa}=1.76,\, \langle\tilde r\rangle_{\rm africa}=0.52$, indicating that the classical dynamics is chaotic~\cite{Chavda2014}. 
Below the upper vHS we obtained for the Africa-shaped DB very good agreement with the GOE. For the rectangular DB an agreement of the ratio distribution with the Poissonian statistics is only achieved after omitting $\simeq 20$ eigenfrequencies $f_i$ closest to the VHS. Then we obtain  $\langle r\rangle_{\rm rect}=2.19 ,\, \langle\tilde r\rangle_{\rm rect}=0.44$, otherwise $\langle r\rangle_{\rm rect}=1.86 ,\, \langle\tilde r\rangle_{\rm rect}=0.49$, in accordance with our observation that the ratio distribution is closer to Poisson and to GOE, respectively. This is again attributed to nongeneric contributions visible in the wavefunction structure. Examples are shown in the middle panels of Fig.~\ref{fig2b}. In both DBs they are localized along zigzag edges within the hexagonal void structure. In the rectangular DB they are scarred along classical trajectories corresponding to particles that bounce back and forth at the two shorter sides of the rectangle. In all considered cases, the ($k$=1)-overlapping ratio distributions obtained by including (green) and omitting (red) nongeneric contributions, are indistinguishable. Finally we note that the highly precise data from the superconducting billiards were a prerequisite for a first rigorous test of the ratio distribution method.  

{\it Conclusions.}---
We measured with unprecedented accuracy the eigenfrequencies of two DBs, that had the shapes of a classically regular rectangular billiard and a chaotic Africa-shaped one, respectively. Like in a graphene sheet or, generally, in artificial graphene, the smooth part of the DOS of the DBs has a complicated structure and exhibits sharp peaks at the vHSs. There, the electronic properties of graphene and likewise, the wavefunction structures calculated with the TBM, change rapidly with increasing frequency. We demonstrated that nevertheless the spectral properties of the DBs are only determined by their shapes. In the vicinity of the VHSs the commonly used statistical measures are not applicable. We, therefore, analyzed the ratio distribution and the ($k$=1)-overlapping ratio distribution and demonstrated that they provide useful statstical measures for the spectral properties. 

This work was supported by the Deutsche Forschungsgemeinschaft (DFG) within the Collaborative Research Center 634. One of us (C.B.) is grateful for the hospitality during an internship at the Institute of Nuclear Physics of the Technical University of Darmstadt.
%
\end{document}